# Neural Speech Tracking in a Virtual Acoustic Environment: Audio-Visual Benefit for Unscripted Continuous Speech

Daeglau, Mareike[1], Otten, Jürgen[2], Grimm, Giso[2], Mirkovic, Bojana[1], Hohmann, Volker[2] & Debener, Stefan[1]

[1] Neuropsychology Lab, Department of Psychology, Carl von Ossietzky University of Oldenburg, Germany
[2] Department of Medical Physics and Acoustics, Carl von Ossietzky University of Oldenburg, Germany

## 1 Abstract

The audio-visual benefit in speech perception—where congruent visual input enhances auditory processing—is well-documented across age groups, particularly in challenging listening conditions and among individuals with varying hearing abilities. However, most studies rely on highly controlled laboratory environments with scripted stimuli. Here, we examine the audio-visual benefit using unscripted, natural speech from untrained speakers within a virtual acoustic environment. Using electroencephalography (EEG) and cortical speech tracking, we assessed neural responses across audio-visual, audio-only, visual-only, and masked-lip conditions to isolate the role of lip movements. Additionally, we analysed individual differences in acoustic and visual features of the speakers, including pitch, jitter, and lip-openness, to explore their influence on the audio-visual speech tracking benefit. Results showed a significant audio-visual enhancement in speech tracking with background noise, with the masked-lip condition performing similarly to the audio-only condition, emphasizing the importance of lip movements in adverse listening situations. Our findings reveal the feasibility of cortical speech tracking with naturalistic stimuli and underscore the impact of individual speaker characteristics on audio-visual integration in real-world listening contexts.

## 2 Introduction

Auditory attention decoding (AAD) has traditionally aimed to distinguish between target and non-target speakers in environments with competing voices, capturing selective attention mechanisms in complex auditory scenes. Significant strides have been made in this field by decoding the speaker to whom a listener is attending, based on the brain's response to multiple simultaneous speakers (Ding & Simon, 2014; Luo & Poeppel, 2007; Mirkovic et al., 2015). AAD studies typically rely on controlled, multi-speaker environments, often using professional speakers and scripted speech for consistency and precision (Holtze et al., 2023; Jaeger et al., 2020; Mirkovic et al., 2016). Although these studies have been foundational, their reliance on controlled settings presents challenges for generalizing findings to more naturalistic auditory environments.

The AAD approach has been extended to explore the impact of visual cues—such as lip movements and facial expressions—on selective attention, especially in noisy settings (Chandrasekaran et al., 2009; Fu et al., 2019). Visual input can enhance speech comprehension by providing congruent cues that aid auditory processing, particularly when auditory signals are degraded. Conversely, incongruent visual cues can create perceptual illusions, demonstrated by the McGurk effect, where mismatched audio and visual inputs can lead to the perception of a novel sound (Jiang & Bernstein, 2011; Mcgurk & Macdonald, 1976; Stropahl & Debener, 2017). This phenomenon underscores the intricate interplay between auditory and visual processing. However, audio-visual fusion seems to vary strongly between different speakers, different audio-visual stimulus combinations, and between participants (Mallick et al., 2015; Stropahl et al., 2017; Stropahl & Debener, 2017).

Most AAD studies are based on neural tracking procedures, which can be used to study how well brain activity captures continuous speech stream fluctuations (Crosse et al., 2015; Luo & Poeppel, 2007; Puschmann et al., 2019). Neural tracking is especially valuable for studying the neural dynamics of speech processing in naturalistic environments, where congruent multi-modal cues, such as lip movements, enhance speech comprehension without the complexity of competing voices.

Traditional context factors such as background noise, speaker position or varying hearing abilities have been thoroughly investigated in AAD and neural speech tracking studies (Geirnaert et al., 2021; Rosenkranz et al., 2021; Wang et al., 2023; Zion Golumbic et al., 2012). However, other factors, such as the likeability of the speaker or specific speech features of the speaking person, may contribute to how well a speech signal is followed by a listener (Wiedenmann et al., 2023). Research findings in the context of advertising or expert testimony suggest that likeability drives attention, meaning that more likeable people capture greater attention independent of their actions (Fam & Waller, 2006; Younan & Martire, 2021). Likeability, as a socio-emotional factor, may influence listener engagement and attention, potentially modulating speech tracking (Farley, 2008; Li et al., 2023).

Similarly, characteristics like articulation clarity, pitch range, or speech rhythm could contribute to individual differences in neural tracking efficacy. These rather unexplored context factors are particularly relevant for understanding real-world communication, where socio-emotional dynamics and individual speaker traits naturally interact with auditory processing (Bachmann et al., 2021; Etard & Reichenbach, 2019; Peelle & Davis, 2012).

To date, it remains poorly understood which speaker characteristics contribute to effective neural tracking. Most studies in the field utilize professional speakers with precise articulation, creating a controlled foundation for understanding neural tracking mechanisms (Crosse et al., 2015; Jaeger et al., 2020). Real-world listening, however, typically involves understanding untrained speakers, whose articulation, pitch, and spontaneity can vary widely. The influence of individual differences in vocal characteristics on neural tracking efficacy may be of particular relevance when audio-visual cues come into play (Vanthornhout et al., 2018). For example, the emotional expressiveness, facial dynamics, and speech fluency of individual speakers may interact with neural tracking and comprehension in ways that are not yet fully understood (Scherer et al., 2019; Tomar et al., 2024).

Previous studies have integrated visual cues into neural speech tracking to determine how congruent visual information, like lip movements, enhances comprehension in dynamic, noisy contexts (Crosse, Liberto, et al., 2016; Park et al., 2016). These results underscore the powerful role of visual-auditory integration in enhancing speech comprehension under challenging listening conditions. However, the interplay between speech content, speaker characteristics, and listener preferences or biases warrants further exploration.

In this study, we examined how speaker-specific characteristics, such as articulation, pitch, and visual expressiveness, influence neural speech tracking in single-speaker, naturalistic audio-visual scenarios. By incorporating diverse speaker profiles and realistic listening contexts, we aim to shed light on the interplay of individual speaker traits and contextual factors in shaping speech processing. We hypothesized that audio-visual (AV) conditions would yield a benefit in neural speech tracking, reflected by larger envelope tracking in AV compared to audio-only (A) stimuli across individual speakers. Additionally, we explored whether individual differences between speakers, characterized by various speech features, would influence the magnitude of A and AV speech tracking and the AV benefit. By linking these speaker-specific traits to neural responses, this study aims to address the gap in understanding how individual speaker characteristics modulate speech processing in naturalistic, single-speaker scenarios.

## 3 Methods

### 3.1 Participants

Twenty normal hearing participants were recruited for the study. Data from two participants were incomplete owing to technical difficulties and were therefore excluded from further processing. Participants' ages ranged from 22 to 35 years (M:26 years; 13f, 5m). The inclusion criteria were self-reported normal hearing, normal or corrected-to-normal vision, no previous or current neurological or psychological disorders, and native German skills. Participants completed questionnaires covering demographic information and general health assessments and gave written informed consent. The study protocol was approved by the Commission for Research Impact Assessment and Ethics of the University of Oldenburg.

## 3.2 Apparatus

Participants were seated in the centre of a cylindrical projection screen, which had a radius of 1.74 m and a height of 2 m (Hohmann et al., 2020). A circular array of 16 active loudspeakers (Genelec 8020C) was positioned behind a screen. Behind the loudspeakers, which were positioned at ear level, was a heavy black curtain to reduce reflections and ambient light, and to provide acoustic treatment at mid and high frequencies. The video image was projected with a single ultra-short throw projector (NEC U321H) at a resolution of 1920 × 1080 pixels at 60 fps. The screen warping was processed in the graphics card (Nvidia Quadro M5000), and the field of view was 120 degrees. Due to the screen warping, the effective pixel density varied across the projection and was lowest in the centre, so the projected video was shifted to one side to achieve the highest possible pixel density. The Toolbox for Acoustic Scene Creation and Rendering (TASCAR) (Grimm et al., 2019) was used for audio playback, control of the virtual acoustic environment in the lab, data logging of all sensors, and experimental control. The videos were embedded in a simple 2D virtual visual environment rendered using the Blender game engine (version 2.79c). The content of the game engine (selection of videos, timing of video playback, position of virtual objects) was controlled by the acoustic engine TASCAR.

## 3.3 Stimuli

For this study, 18 videos, each comprising one of six different speakers (2m; 1d; 3f) were taken from a set of pre-recorded audio-visual stimuli (2023; Wiedenmann et al., 2023). Speakers sat in front of a dark grey background, showing their head and upper body up to their shoulders centred in frame. Speakers talked continuously at their natural pace in standard German about self-selected content right into the camera, but with natural movements and glances wandering occasionally. The duration of the videos varied between 180 s and 600 s, cut down from longer recording sessions. The videos were recorded using a Canon EOS 700D with a resolution of 1920 × 1080 pixels at 25 fps. The corresponding audio was recorded with a cardioid microphone (Neuman KM184) at approximately 0.7 m, using an RME Micstasy preamplifier and AD-converter with 48 kHz sampling rate. Speakers wore open earphones (https://batandcat.com/portable-hearinglaboratory-phl.html), in which half of the recordings played babble background noise at a sound pressure level (SPL) of 65 dB unweighted. Video editing and audio and video synchronization were performed using DavinciResolve (Version17). Videos and audio were processed using FFMPEG (http://www.ffmpeg.org).

Each video was cut into consecutive 30-second segments and the following conditions were prepared: audio-visual (AV), audio-only (A), visual-only (V) and masked-lips (ML). In the AV conditions the speaker was presented with the corresponding audio; in the audio-only conditions, the audio was presented alongside a video of a grey-background; in the V-conditions, only the video of the real speaker was shown while the audio was muted; for the ML conditions, the lips of the speaker were overlaid with a light blue horizontal bar, while the corresponding audio was played unaltered. The order of these conditions was pseudo-randomized within the 18 videos but kept constant across participants, changing every 30 s. However, the order of presentation of the 18 videos was randomized across participants. Half of the conditions were block wise presented with background noise (65 dB SCT), while the other half was presented in quiet conditions. Participants were instructed to pay attention to the speakers and the content of the stories. After each experimental session, the participants were asked to rate the likeability of the speakers and how well they were able to follow their stories on a five-point Likert scale. Additionally, questions about the story content were asked in a multiple-choice format. Due to the total duration of 85 min of story content, the experiment was split into two sessions to avoid exhausting participants and potentially distorting neural speech tracking. The period between both sessions varied between one and 14 days.

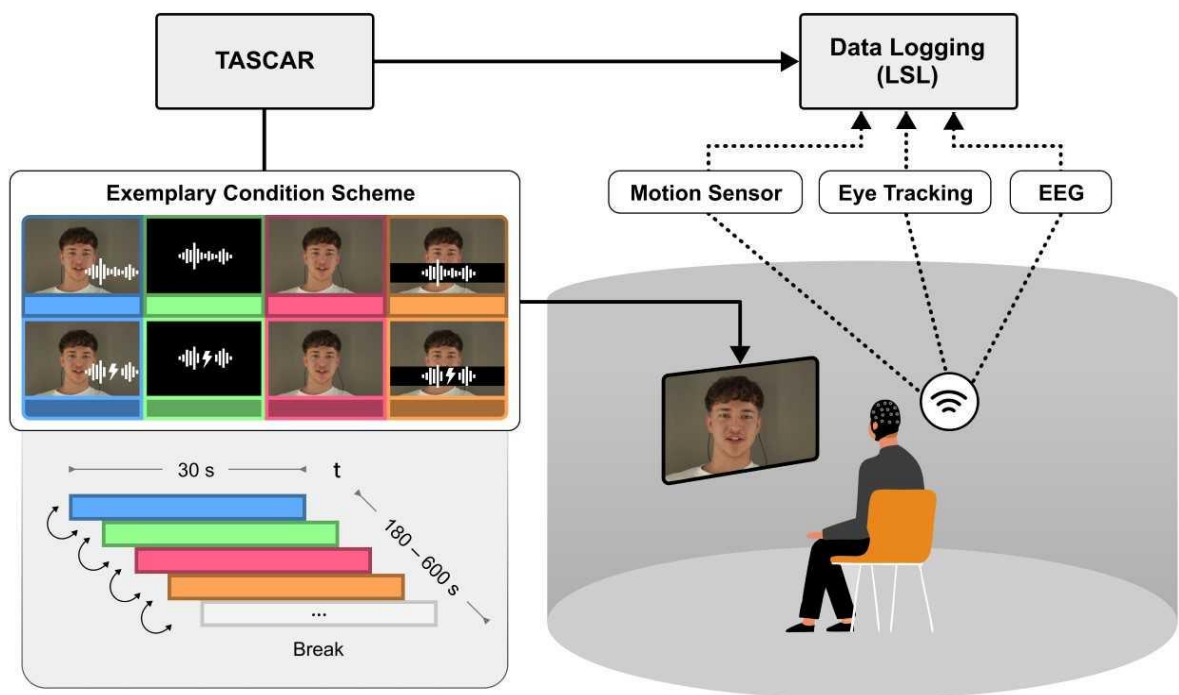

Figure 1: Experimental set-up including data logging. Using TASCAR VE, 18 participants were presented audio, visual or congruent combined audio-visual scenes comprising one out of six speakers at a time telling unscripted stories. Listening difficulty was manipulated by comparing

speech with and without babble noise. Data logging was achieved wirelessly via Bluetooth and Wi-Fi.

3.4 EEG data acquisition and preprocessing

EEG data were acquired using a wireless, head-mounted 24-channel EEG system (SMARTING, mBrainTrain, Belgrade, Serbia). The system features a sampling rate of 500 Hz, a resolution of 24 bits, and a bandwidth from DC to 250 Hz. EEG data were collected from 24 scalp sites using sintered Ag/AgCl electrodes with FCz as the ground and AFz as the reference (Easycap, Herrsching, Germany). The electrode sites were prepared using 70% alcohol and an abrasive electrolyte gel (Abralyt HiCl, Easycap GmbH, Germany). The electrode impedances were maintained below 10 kΩ and tested before data acquisition. The EEG signal as well as other data (i.e., eye tracking, head movements; not investigated here) were wirelessly transmitted to a PC via Bluetooth and synchronized and recorded using the lab streaming layer protocol (Kothe et al., 2024) and saved into an .xdf file. Additional data recording was performed using TASCAR to .mat files (eye and head tracking, not analysed here).

For offline analysis, EEGLAB (Delorme & Makeig, 2004) and MATLAB (R2024a, MathWorks Inc., Marick, MA, USA) were used. Identification of improbable channels was conducted using the EEGLAB extension trimOutlier with an upper and lower boundary of two standard deviations of the mean standard deviation across all channels. Channels that exceeded this threshold were excluded. A copy of the EEG data was first low-pass filtered at 40 Hz (finite impulse response (FIR), Hamming window, filter order 166), downsampled to 250 Hz, and subsequently high-pass filtered at 1 Hz (FIR, Hamming window, filter order 414; filters integrated into EEGLAB, version 1.6.2). Afterwards, data were segmented into consecutive 1-s epochs and segments containing artifacts were removed (EEGLAB functions pop eegthresh.m, +-80uV; pop rejkurt.m, SD = 3). The remaining data were submitted to extended Infomax ICA. The unmixing matrix obtained from this procedure was applied to the original unfiltered EEG dataset to select and reject components representing stereotypical artifacts. Components reflecting eye, muscle, and heart activity were identified using ICLabel (Pion-Tonachini et al., 2019). Components flagged and identified as artifacts were removed from further analysis. Artifact-corrected EEG data were low-pass filtered with a FIR filter and a cut-off frequency of 30 Hz (hann window, filter order 220, Fs = 500 Hz), and subsequently high-pass filtered with a FIR filter and a cut-off frequency of 0.3 Hz (hann window, filter order 500, Fs = 500 Hz). After the data were re-referenced to the common average and corrupted channels were replaced by spherical interpolation, the data were resampled to 64 Hz (to reduce the computational demand for the envelope reconstruction) and cut into 30-sec epochs

(matching the presentation of conditions in the experiment). Pre-processed EEG data were further processed using the mTRF toolbox (Crosse, Di Liberto, et al., 2016).

### 3.5 Audio pre-processing and speech envelope reconstruction

A broadband audio envelope was extracted as follows: Each audio track was z-normalized and bandpass filtered into 128 logarithmically-spaced frequency bands between 100 and 6500 Hz, using a gamma tone filter bank (Herzke & Hohmann, 2007; Hohmann, 2002). The 100–6500 Hz range was chosen based on previous research suggesting a high temporal coherence between visual features and speech envelope within this frequency range (Chandrasekaran et al., 2009; Crosse et al., 2015). Hilbert transformation was used to compute the signal envelope within each of 128 frequency bands. The broadband envelope was then obtained by averaging the absolute Hilbert values across all bands. The broadband envelope was low-pass filtered at 30 Hz using a 3rd-order Butterworth filter and subsequently down-sampled to 64 Hz for further processing. The mTRF toolbox (Crosse, Di Liberto, et al., 2016), was used to reconstruct the broadband envelope utilizing the presented speech signals and the EEG data. This approach is based on multivariate linear regression to obtain a linear mapping between the EEG sensor data and the broadband speech envelope. The determination of the ridge parameter λ was achieved through an optimization process involving a search grid and a leave-one-out cross-validation procedure to minimize the mean-squared error associated with the regression. The range of values within the search grid encompassed magnitudes such as $10^{-2}$, $10^{-1}$, ..., $10^4$, $5 \times 10^4$, $10^5$, …, $10^9$. To ensure the generalizability of the established relationship between speech input and neural response, a subject-level leave-one-out cross-validation strategy was executed. This entailed the reconstruction of the speech envelope for a specific trial using the mean regression weights derived from all other trials within the same experimental condition and at the same temporal lag. The reliability of the reconstruction was quantified by computing Pearson's correlation coefficient between the reconstructed and original speech envelopes. For statistical treatment, the correlation coefficients were subjected to Fisher's z-transformation to achieve normality and were subsequently averaged across trials. For an initial exploration of the temporal dynamics of speech envelope tracking, individual lag models, characterized by 24 regressors corresponding to each EEG channel, were computed for every trial across 33 discrete time lags spanning from stimulus presentation to EEG signal acquisition, covering a temporal range of 0 to 500 ms. This analysis yielded a time course, from which the time window of interest was discerned (200-325 ms). For further analyses of audio-visual enhancements, multi-lag models containing 24 x N(lags) regressors were computed for each of these time windows and all trials (Puschmann et al., 2017, 2019). For further statistical evaluation, r values were normalized using MATLAB's atanh-function ($r_z$).

## 2.6 Exploratory Analyses

### 2.6.1. Extraction of Acoustic Features

To analyze oscillatory components in the audio data, the frequency spectrum was divided into four bands: envelope-range (0.3–30 Hz), low-range (30–300 Hz), mid-range (300–1000 Hz), and high-range (1000–4500 Hz). This division allowed for a detailed examination of low-frequency elements associated with prosody and high-frequency components characteristic of speech. MATLAB and the FieldTrip toolbox were employed to implement the multitaper method (mtmfft) for frequency-domain analysis, well-suited for the relatively short (30-second) audio segments in this study. Each audio segment was transformed into a power spectrum under specific configurations. Frequency smoothing was set at 0.5 Hz (cfg_tapsmofrq = 0.5), balancing resolution and noise reduction across frequency bands. The analysis was limited to a frequency range of 0.3 to 4500 Hz (cfg.foilim) to exclude non-speech-relevant frequencies. To isolate oscillatory components in the data a division approach was employed. The original power spectrum was normalized by dividing it by the fractal component, reducing the influence of non-oscillatory noise (cfg.operation = 'x2./x1'). For each speaker, periodic power within each frequency band was summed and normalized by the segment duration, resulting in an average periodic power per band, which was stored for further analysis, respectively, FreqRsum<30 (envelope-range), FreqRsum<300 (low-range), FreqRsum<1k (mid-range), and FreqRsum<4.5k (high-range), indicating the amount of periodic proportions for each speaker. Additionally, a set of 16 acoustic features was extracted from each 30-second audio segment to capture essential elements of vocal dynamics and quality with Praat (Boersma, Paul & Weenink, David, 2024) using the in-build voice report metrics. These features included *Pitch Metrics* (meanPitch, medianPitch, sdPitch, minPitch, maxPitch), *Jitter Metrics* (jitter_loc, jitter_loc_abs, jitter_rap, jitter_ppq5), *Shimmer Metrics* (shimmer_loc, shimmer_loc_dB, shimmer_apq3, shimmer_apq5, shimmer_apq11), *Noise-to-Harmonic Ratio (NHR)* (mean_nhr) and *Intensity* (min_intensity). For each speaker, each feature was averaged across segments to reduce inter-segment variability, providing a robust profile for inter-speaker comparison.

### 2.6.2 Extraction of Visual Features

To consider the multimodal nature of our stimuli, two visual features were extracted from each video segment using a custom Python-based image processing script. The script specifically targeted *Lip Openness* (representing articulatory movements associated with speech) and *Lip Brightness* (capturing the visual clarity and lighting conditions of each video segment). Using

OpenCV, the Python script processed video data to compute average values for each visual feature over the segment duration.

### 2.6.3 Feature processing

After feature extraction, including frequency-based, acoustic, and visual data, features were normalized from 0 to 1 using MATLAB's normalize function, facilitating comparability across features with different scales. To refine the feature set and emphasize the most informative variables, features that were correlated above an r-value of 0.8 were removed. Each of the remaining feature's correlations with the average condition values obtained prior (see section 2.5) was assessed.

## 2 Results

### 3.1 Neural speech tracking across conditions

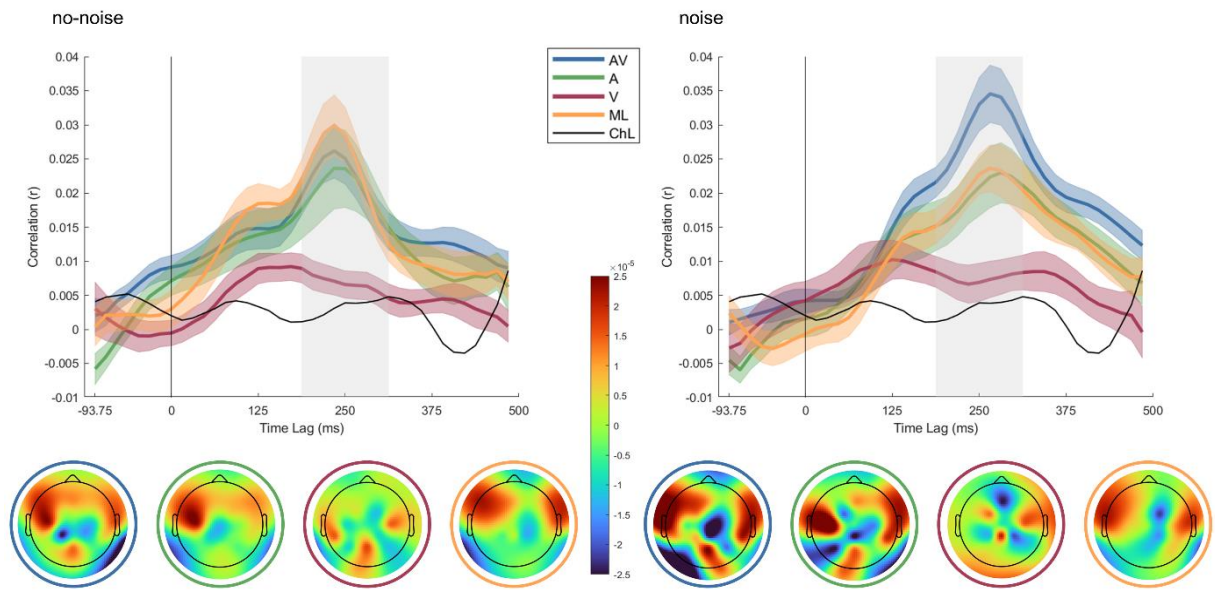

Fig. 2 Temporal dynamics of cortical speech tracking across different audio-visual conditions in noise and no-noise environments and decoder weight topographies. In the upper panels, line plots display the time-lagged correlation (r) between the neural response and the speech envelope for four conditions: audio-visual (AV) in blue, audio-only (A) in green, visual-only (V) in red, and masked-lips in orange, in both noise (right) and no-noise (left) conditions. The chance level (ChL), computed via random permutation between EEG and audio data, is depicted in black. The x-axis represents the time lag (in ms) relative to the trial onset, with positive values indicating the delayed neural response, while the y-axis shows the correlation coefficient (r).

The AV condition exhibits the highest correlation in the presence of noise, peaking around 250 ms, whereas the ML and A conditions yield lower, thus comparable correlations. The V condition correlates lowest, but in most time lags above the chance level. The lower panels display topographic maps showing the decoder weight distribution of neural responses across the scalp for each condition in both no-noise and noise contexts. Each map represents the condition specific decoder weights, with colour gradients indicating the strength and direction of the weighting. In the noise condition, AV and A show distinct patterns in frontal and temporal regions, suggesting enhanced neural tracking when both audio and visual cues are present. In the no-noise condition, the spatial response patterns are more evenly distributed, with AV and A conditions still demonstrating more pronounced activations than V or ML.

Fig. 2 depicts the speech envelope reconstruction accuracy $r_z$ for each listening condition as a function of the relative time lag between auditory input and EEG response. The time window of interest (i.e., 200–325 ms; indicated in grey) was defined based on the group-level peaks of envelope reconstruction accuracy in all conditions. To investigate expected differences in $r_z$ between conditions, we performed a 2x4 repeated measures ANOVA with $r_z$ as the dependent variable and two within-subject factors: background noise (two levels: noise, no-noise) and audio-visual effect (four levels: congruent audio-visual, visual-only, audio-only and masked-lips). ANOVA results indicated a significant main effect for audio-visual effect ($F_{1.72,29.3}$ = 16.95, $p$ = < .001, η2 = 0.36), and a significant interaction effect for background noise x audio-visual effect ($F_{3,51}$ = 7.51, $p$ = .08, η2 = 0.03) but no significant main effect for background noise ($F_{1,17}$ = 0.06, $p$ = < .001, η2 = 0.002). Planned post hoc paired t-tests revealed significant differences between AV and A in noise ($t_{(17)}$ = 3.91, $p$ = .001, d = 0.87) and AV and ML in noise ($t_{(17)}$ = 3.71, $p$ = .002, d = 0.92), with AV being more pronounced then A or ML, but not between A and ML in noise ($t_{(17)}$ = -0.37, $p$ = .72, d = 0.09). Further, for the no-noise conditions, no significant differences between AV and A ($t_{(17)}$ = 0.38, $p$ = .071, d = 0.09), AV and ML ($t_{(17)}$ = -1.01, $p$ = .33, d = -0.24), and A and ML ($t_{(17)}$ = -1.11, $p$ = .28, d = -0.26). Additionally, AV in noise was significantly higher than AV no-noise ($t_{(17)}$ = 2.99, $p$ = .008, d = 0.7). P-values were corrected for multiple comparisons over seven tests using Holm-Bonferroni (Holm, 1979).

### 3.2 Exploratory results
#### 3.2.1 Cortical speech tracking for conditions AV, A and ML for each speaker

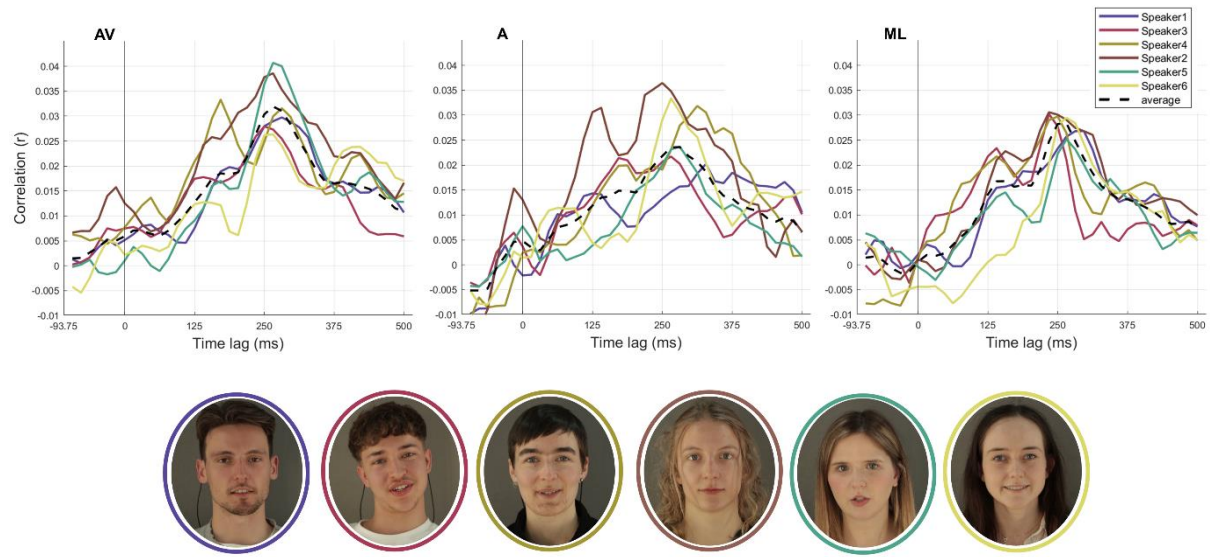

Fig. 3 Cortical speech tracking for AV, A and ML conditions for all six speakers. The top row illustrates the time-lagged correlations ($r_z$) between audio envelopes and neural responses for each of the six speakers (speaker1, speaker3, speaker4, speaker2, speaker5, speaker6). Conditions include audio-visual (AV) on the left, auditory-only (A) in the centre, and masked-lips (ML) on the right. Individual speaker data are represented as coloured solid lines, with colours corresponding to each speaker (speaker1: purple, speaker3: red, speaker4: yellow-green, speaker2: brown, speaker5: teal, speaker6: yellow). The black dashed line represents the average envelope across speakers. The bottom row displays photographs of each speaker, bordered in colours corresponding to their respective line plots in the top row. Conditions are not separated for noise and no-noise conditions due to the limited number of trials for each individual speaker.

On average, participants rated all six speakers comparably high in likeability on a scale from 0 - 5 (Mean: 3.79 ± 0.02, range: 3.47 - 3.94). Pictures of all speakers as well as individual neural time courses for conditions AV, A and ML and respective averages are depicted in Figure 3. Time courses show similar patterns across speakers over the time window of interest (200-325 ms) but are diverging in an earlier time window (~125 ms). In the A condition, speaker2 shows the most pronounced response pattern.

### 3.2.2 Relationships between auditory envelopes, audiovisual conditions, and acoustic features

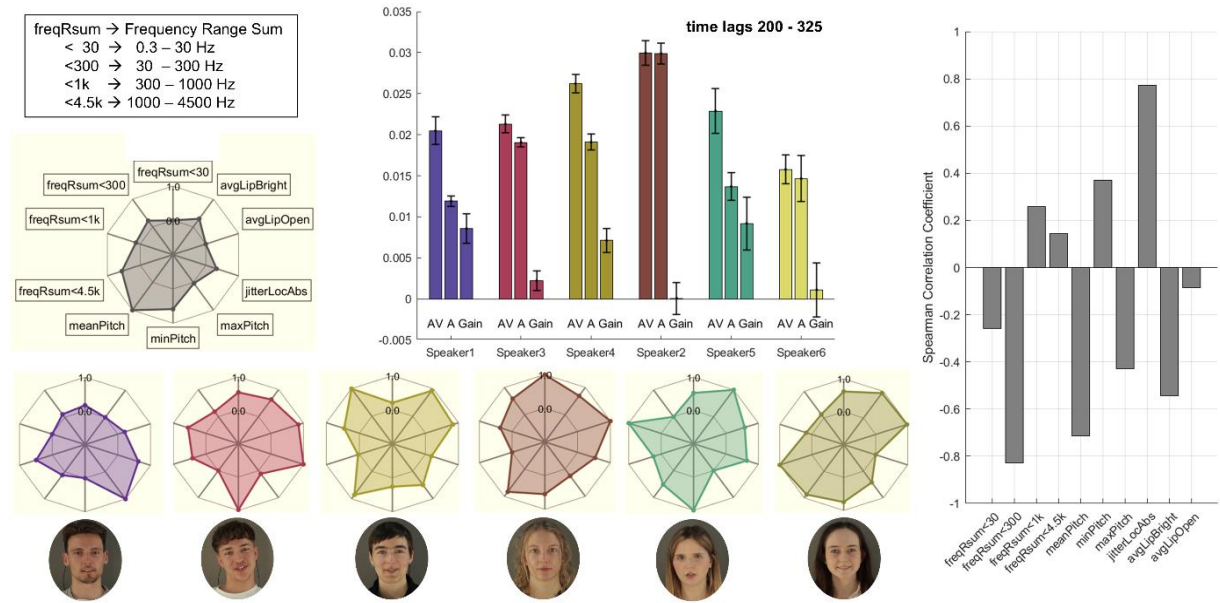

Fig. 4: Relationships between auditory envelopes, audiovisual conditions, and acoustic features.
(Top Row): The barplot illustrates the mean correlation (r$_z$) across time lags 200–325 ms for audiovisual (AV), audio-only (A), and audiovisual gain (AV-A; BF) conditions for each speaker (speaker1, speaker3, speaker4, speaker2, speaker5, speaker6). Error bars represent the standard error of the mean.
(Middle Row): Radar plots depict acoustic feature distributions per speaker, including frequency range sums (freqRsum <30 Hz, <300 Hz, <1k Hz, <4.5k Hz), pitch features (mean, min, max), jitter (jitterLocAbs), and lip-based brightness and openness averages (avgLipBright, avgLipOpen). Each radar plot highlights inter-speaker variability across the selected features. (Bottom Row): Portraits of the six speakers visually align with their corresponding radar plots and barplots, facilitating a direct comparison of individual acoustic feature profiles and correlation patterns.
(Right Panel): Spearman correlation coefficients display the relationship between selected acoustic features and the audiovisual benefit (AV-A), highlighting feature-specific contributions to the benefit.

The analysis revealed consistent and distinct patterns across auditory and visual features, time-lagged correlations, and gain-specific conditions. Summed power within predefined frequency bands (freqRsum<30, freqRsum<300, freqRsum<1k, freqRsum<4.5k) demonstrated variability across speakers. Notably, speaker2 showed the highest summed power in the lower frequencies, speaker4 in the low-range, speaker5 in the mid-range and speaker6 in the high-range, respectively. MaxPitch is highest for speaker1, whereas speaker3 and speaker2 share highest minPitch. Condition-specific analysis focusing on audio-visual (AV) and audio-only (A) correlations within the critical 200–325 ms time window identified a general increase in AV conditions for most speakers. Speaker2 induced the highest AV and A correlations, simultaneously also the smallest audiovisual gain, as reflected in the bar plot. Visual features derived from lip brightness (avgLipBright) and openness (avgLipOpen) displayed notable differences. All speakers, except for speaker1, exhibited higher values for these visual features, suggesting more expressive visual characteristics for these speakers. Radar plot representations further illustrated the unique multimodal profiles, capturing variability across frequency sums, pitch, jitter measures, and visual parameters.

Correlation analysis between barplot condition averages (AV, A, BF) and selected features from the radar plots indicated potential relationships. For audiovisual benefit (Gain), the strongest positive correlation was observed with freqRsum<30 (r ≈ 0.85), suggesting that low-frequency power may play a prominent role in audio-visual benefit. Additional correlations emerged between audiovisual benefit and meanPitch and avgLipOpen, indicating modest but significant associations. Please note that none of the correlations reached significance (only comprising six datapoints each) and serve purely for descriptive purposes.

## 4 Discussion

This study investigated whether young, normal-hearing individuals benefit from congruent facial cues of speakers, when listening to unscripted, natural speech in both quiet and noisy environments. Our results demonstrate that congruent audio-visual input enhances neural speech tracking in noise, with significantly higher correlations ($r_z$) in the audio-visual condition compared to audio-only and masked-lips conditions. These findings support the well-established notion that visual input facilitates speech perception, particularly in challenging listening environments (Peelle & Sommers, 2015; Sumby & Pollack, 2005), even in participants with (self-reported) normal hearing abilities.

### 4.1 Audio-Visual Enhancement of Neural Speech Tracking

Our findings reveal a clear audio-visual benefit for speech envelope tracking, particularly in noisy environments. This aligns with previous studies showing that visual cues enhance auditory processing when the speech signal is degraded (Peelle & Sommers, 2015; Zion Golumbic et al., 2013). In our results, the AV condition consistently induced higher correlation values compared to both the audio-only and masked-lips conditions in noise. Importantly, AV tracking peaked at around 250 ms, a time window consistent with cortical auditory-visual integration processes. This supports the idea that visible lip movements help align auditory cortical oscillations with the speech envelope (Arnal & Giraud, 2012; Schroeder et al., 2008). The lack of significant differences between AV and A in the no-noise condition suggests that visual cues primarily become beneficial when the auditory input is compromised, as noted in earlier work (Sumby & Pollack, 2005). In contrast, the absence of a significant difference between ML and A in noise highlights the specific role of visible lip movements in driving the AV benefit. This finding underscores that visual articulation cues are central to the AV advantage in neural speech tracking models.

Interestingly, in the no-noise condition, the masked-lips condition showed, on a descriptive level, even higher neural tracking than the audio-visual condition. Several studies (Rahne et al., 2021; Sönnichsen et al., 2022) conclude that face masks reduce speech perception and increase listening effort in different noise signals even in normal hearing participants. A relevant contribution to this effect, masked-induced auditory deterioration, was not included in our study. We speculate that listeners may have adpated to auditory-only communication during the COVID-19 pandemic, when face masks frequently obscured visual cues like lip movements and caused auditory degradation. Research suggests that prolonged exposure to masked faces can lead to increased reliance on auditory processing and reduced dependence on visual input (Saunders et al., 2021). In the no-noise condition, where the auditory signal was clear and unaltered, participants may have defaulted to auditory-only strategies, ignoring the incongruent or incomplete visual cues in the ML condition. This could have reduced cognitive load, allowing for more efficient speech envelope tracking compared to AV, where lip movements might introduce redundant or misaligned visual information (Yi et al., 2021).

Our study's focus on unscripted, naturally told stories adds ecological validity by resembling real-world listening conditions, where continuous speech provides contextual richness. This approach more effectively enhances neural speech tracking compared to isolated words or sentences, as previously demonstrated (Gross et al., 2013). By contrast, studies relying on highly controlled stimuli may miss the natural dynamics of conversation. While our findings broadly align with prior research, they diverge from a study reporting no AV benefit in single-speaker contexts (O'Sullivan et al., 2013). This discrepancy could arise from differences in stimulus duration or the specific envelope tracking methods used.

### 4.2 Speaker-Specific Differences in Neural Speech Tracking

We observed considerable variability in neural speech tracking across speakers, especially in the A and AV conditions. For example, speaker2 exhibited the highest overall tracking correlations in both conditions but had the smallest audio-visual gain. In contrast, speakers like speaker4 and speaker6 demonstrated pronounced AV benefits, suggesting that individual speaker characteristics influence how effectively visual and auditory inputs integrate. Specific acoustic traits, such as speaker3's stronger mid-range spectral power or speaker6's higher-range power, may influence their ability to engage neural tracking. Pitch variability also played a role: speaker6's higher mean pitch and speaker3's wider pitch range likely contributed to distinct neural representations that aid speaker differentiation (Bidelman & Howell, 2016). Brodbeck and Simon (2022) demonstrated that voice pitch variability significantly modulates cortical neural tracking, particularly under conditions requiring selective attention to distinct speakers. These results highlight that speaker-specific traits—including frequency content and articulation variability—shape the dynamics of neural speech tracking.

Despite this variability, participants rated all speakers similarly on likeability, suggesting that subjective perception was not strongly influenced by individual acoustic or visual differences. This is consistent with prior findings showing stable likeability ratings across speaker profiles (Zuckerman & Driver, 1989). The observed speaker-specific differences emphasize the importance of accounting for individual multimodal profiles when studying neural speech tracking. This variability is particularly relevant for real-world scenarios where listeners engage with speakers of diverse expressiveness and acoustic profiles.

### 4.3 Relationships Between Speech Features and Neural Correlations

Our exploratory analysis uncovered relationships between speaker-specific acoustic features and audiovisual benefit (AV-A). Low-frequency spectral power (freqRsum<30) was most strongly linked to AV benefit, suggesting that low-frequency components are particularly crucial for driving neural speech tracking. This finding aligns with earlier research emphasizing the importance of low-frequency energy for neural speech tracking (Ding & Simon, 2014; Luo & Poeppel, 2007). Jitter measures, such as jitter_loc_abs, also emerged as relevant predictors, indicating that micro-level vocal irregularities could enhance perceptual salience under noisy conditions (Eadie & Doyle, 2005; Oganian et al., 2023). These vocal perturbations, including jitter and shimmer, are known to modulate speech clarity and contribute to perceptual robustness in degraded auditory environments (Smiljanic & Gilbert, 2017). While exploratory, our findings hint at a complex interplay between acoustic clarity and neural tracking dynamics. On the visual side, articulatory expressiveness as measured by lip openness showed modest positive associations with AV benefit. This suggests that more pronounced visual articulation aids the brain in aligning auditory and visual information. Overall, these results point to a combined contribution of acoustic spectral features and visual articulation cues in driving audiovisual integration.

### 4.4 Implications for Multimodal Speech Processing

Our findings have significant implications for understanding how the brain integrates auditory and visual cues during natural, unscripted speech. In addition to studies using controlled or scripted stimuli, we show that neural speech tracking is also robust in more ecologically valid listening conditions. The enhanced tracking observed in noisy AV conditions highlights the critical role of visible lip movements in compensating for degraded auditory signals, emphasizing the importance of cross-modal integration in real-world communication.

From an application perspective, these results can inform technologies like hearing aids and brain-computer interfaces. Incorporating speaker-specific acoustic and visual profiles could improve auditory attention decoding models, optimizing neural tracking performance in naturalistic settings (Geirnaert et al., 2021). Understanding how individual speaker traits influence audiovisual integration – and attention - is crucial for developing personalized solutions to enhance real-world speech perception.

**4.5 Limitations**

Several limitations should be noted. First, we included only a the relatively small number of speakers (N = 6), which limits the statistical power of our exploratory analyses. While our descriptive trends provide valuable insights, larger datasets including more speaker variability are needed to confirm the role of specific acoustic and visual features in audiovisual benefit. Second, the influence of noise conditions at the speaker level could not be fully explored due to trial limitations, which restricts a more detailed analysis of how speaker traits interact with noise to shape speech tracking. Finally, while participants rated speakers similarly on likeability, subtle socio-emotional factors or individual listener preferences could still have influenced neural responses. Future studies should consider including subjective biases and emotional expressiveness as additional covariates.

# 5 Summary

In summary, this study highlights the interplay between speaker-specific acoustic and visual attributes and their effect on audio-visual integration and neural speech tracking. These insights have implications for personalized auditory attention models and assistive technologies, emphasizing the need to account for individual variability in natural, unscripted multi-speaker environments. Future research should extend these findings by exploring multimodal integration in diverse populations, including those with hearing impairments, to further enhance predictive models of auditory attention.

# 6 Conflict of interest

None

# 7 Data availability statement

## 8   Ethics statement

The studies involving humans were approved by Kommission für Forschungsfolgenabschätzung und Ethik, University of Oldenburg, Oldenburg, Germany. The studies were conducted in accordance with the local legislation and institutional requirements. The participants provided their written informed consent to participate in this study. Written informed consent was obtained from the individual(s) for the publication of any potentially identifiable images or data included in this article.

## 9   Acknowledgments

This work was supported by the Cluster of Excellence "Hearing4all", funded by the German Research Foundation (Deutsche Forschungsgemeinschaft; under Germany's Excellence Strategy – EXC 2177/1 - Project ID 390895286). Mareike Daeglau and Jürgen Otten were supported by a research grant from the German Research Foundation (Deutsche Forschungsgemeinschaft; SPP 2236 project 444761144). Partly funded by the Deutsche Forschungsgemeinschaft (DFG, German Research Foundation, Project-ID 352015383 – SFB 1330 project B1).

The authors thank the speakers (Aaron Möllhof, Anna Dorina Klaus, Janto Klunder, Mara Wendt-Thorne, Laura Knipper & Jupiter Dunkelgut), Focke Schröder and Ina Cera (technical assistance, video editing) and Kevin Brumme (valuable advice on visual designs). The EEG data was acquired with the help of Jennifer Decker and Emma Wiedenmann.